\documentclass[showpacs,preprintnumbers,english,twocolumn,amsmath,amssymb]{revtex4}
\usepackage[T1]{fontenc}
\usepackage[latin1]{inputenc}
\usepackage{graphicx}
\usepackage{amssymb}
\usepackage{verbatim}
\usepackage{epic,eepic}
\usepackage{babel}
\begin{document}
\title{Influence of the Heisenberg Principle on the Ideal Bose Gas}
\author{ Hua Zheng$^{a,b)}$, Gianluca Giuliani$^{a)}$ and Aldo Bonasera$^{a,c)}$}
\affiliation{
a)Cyclotron Institute, Texas A\&M University, College Station, TX 77843, USA;\\
b)Physics Department, Texas A\&M University, College Station, TX 77843, USA;\\
c)Laboratori Nazionali del Sud, INFN, via Santa Sofia, 62, 95123 Catania, Italy.}
\begin{abstract}
The ideal Bose gas has two major shortcomings: at zero temperature, all the particles 'condense' at zero energy or momentum, thus violating the Heisenberg principle; the second is that the pressure below the critical point is independent of density resulting in zero incompressibility (or infinite isothermal compressibility) which is unphysical. We propose a modification of the ideal Bose gas to take into account the Heisenberg principle. This modification results in a finite (in)compressibility at all temperatures and densities. The main properties of the ideal Bose gas are preserved, i.e. the relation between the critical temperature and density, but the specific heat has a maximum at the critical temperature instead of a discontinuity. Of course interactions are crucial for both cases in order to describe actual physical systems.
\end{abstract}
\maketitle
Elementary particles with integer spins, follow the Bose-Einstein (BE) distribution:
\begin{equation}
f(\varepsilon)=\frac{1}{e^{(\varepsilon-\mu)/T}-1}, \label{beq}
\end{equation}
where $\varepsilon$ is the energy, $\mu$ is the chemical potential and $T$ is the temperature. Because of the factor $(-1)$ in the denominator, the chemical potential must be $\mu \le 0$. When $\mu=0$, a critical temperature $T_0$ \cite{landau, huang, pathria} can be defined such that:
\begin{equation}
T_0 = \frac{3.3125}{g^{2/3}}\frac{\hbar^2}{m}\rho^{2/3}, \label{ct}
\end{equation}
where $g$ is the degeneracy, $m$ is the mass and $\rho$ is the density. For $T<T_0$, some of the particles have energy greater than zero, while the remaining particles 'condense' at $\varepsilon=0$.  Recall that in order for this to be true, the total number of particles must be fixed. Strictly speaking the 'condensed' particles violate the Heisenberg principle or, alternatively, the density of the system must be zero, which means that the distance between particles is infinite. Of course this case is not interesting and we propose a modification in order to take into account the Heisenberg principle. This principle states that any two elementary particles cannot be closer in phase space than the Heisenberg constant $\hbar$. We can restate the principle by imposing that in a volume of radius $R$ and momentum $P_H$ one particle alone can be accommodated \cite{sachie, hua}:
\begin{equation}
\frac{1}{\hbar^3} = \frac{N}{V_R V_{P_H}}=\frac{N}{(\frac{4\pi}{3})^2R^3P_H^3}. \label{hp}
\end{equation}
Eq. (\ref{hp}) defines a 'Heisenberg momentum' given by
\begin{equation}
P_H = (\frac{3\rho}{4\pi})^{\frac{1}{3}}\hbar. \label{ph}
\end{equation}
Notice the similarity between the Heisenberg momentum and the Fermi momentum which can be similarly obtained from Eqs. (\ref{hp}, \ref{ph}) by changing $\hbar\rightarrow h$, and including the $g=2s+1$ factor, where $s$ is the particle spin (similarly for the isospin). Thus the Pauli principle, which holds true for identical fermions, can be thought of as a 'stronger' version of the Heisenberg principle \cite{sachie, hua}. From Eq. (\ref{hp}), we can define a 'Heisenberg energy' as:
\begin{equation}
\varepsilon_H = \frac{P_H^2}{2m}= (\frac{3}{4\pi})^{\frac{2}{3}}\frac{\hbar^2}{2m} \rho^\frac{2}{3}.\label{he41, he42, he43, he44}
\end{equation}

Notice that the Heisenberg energy (similar to the Fermi energy) has the same density dependence of the critical $T_0$,  Eq. (\ref{ct}), but with a smaller coefficient thus resulting in $\varepsilon_H<T_0$. 
In order to take into account the Heisenberg principle we have to impose that no particles violate the Heisenberg principle.  This can be fulfilled by re-writing the BE integrals with the lower limit $\bar\varepsilon_H = \frac{3}{5}\varepsilon_H$ instead of zero, which is our ansatz.  For instance the density is calculated as:
\begin{eqnarray}
\rho&=&\frac{N}{V}\nonumber\\
&=& \frac{g2\pi(2m)^{3/2}}{h^3}\int_{\bar\varepsilon_H}^\infty d\varepsilon \frac{\varepsilon^{1/2}}{e^{(\varepsilon-\mu)/T}-1} \nonumber\\
&=& \frac{g(2\pi mT)^{3/2}}{h^3}h_{3/2}(z, T), \label{density}
\end{eqnarray}
where we define the function $h_n(z, T)=\frac{1}{\Gamma(n)}\int_{\bar\varepsilon_H/T}^\infty dx \frac{x^{n-1}}{z^{-1}e^{x}-1}$ and $z=e^{\mu/T}$. The important consequence is that the chemical potential might be positive and at most:
\begin{equation}
\mu \rightarrow \bar\varepsilon_H \quad\text{for}\quad T\rightarrow 0. \label{mul}
\end{equation}
Eq. (\ref{density}) tells us that such a solution is possible even though, for very small $T$, numerical solutions become unstable, but still feasible as we will show below. 
\begin{figure}
\centering
\includegraphics[width=1.0\columnwidth]{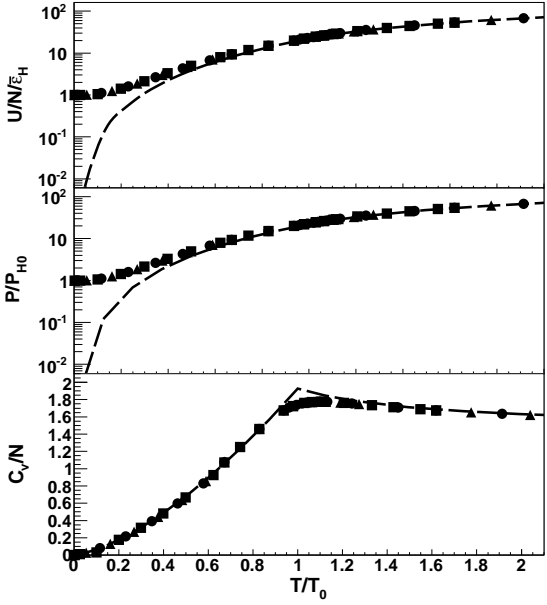}
\caption[]{Energy per particle, pressure (normalized by their corresponding Heisenberg values, Eqs. (\ref{unl}) and (\ref{pt0})), specific heat vs $T/T_0$. The symbols refer to three different densities (0.004, 0.04 and 0.4 in arbitrary units) while the dashed line is the ideal BE gas results \cite{landau, huang, pathria}.}\label{Fig1}
\end{figure}
In order to get convinced that our solution does not violate the Heisenberg principle, we evaluate the energy per particle at zero temperature:
\begin{eqnarray}
\frac{U}{N}\Big|_{T\rightarrow 0}&=&\frac{\int_{\bar\varepsilon_H}^\infty d\varepsilon \varepsilon^{3/2} \frac{1}{e^{(\varepsilon-\mu)/T}-1}}{\int_{\bar\varepsilon_H}^\infty d\varepsilon \varepsilon^{1/2} \frac{1}{e^{(\varepsilon-\mu)/T}-1}}\Big|_{T\rightarrow 0}\nonumber\\
&\approx&\frac{\int_{\bar\varepsilon_H}^\infty d\varepsilon \varepsilon^{3/2} e^{-(\varepsilon-\mu)/T}}{\int_{\bar\varepsilon_H}^\infty d\varepsilon \varepsilon^{1/2} e^{-(\varepsilon-\mu)/T}}\Big|_{T\rightarrow 0}\nonumber\\
&\rightarrow& \bar\varepsilon_H. \label{unl}
\end{eqnarray}
Notice the similarity to the average Fermi energy. At finite temperatures, the energy per particle is given by:
\begin{equation}
\frac{U}{N}=\frac{3}{2}T\frac{h_{5/2}(z, T)}{h_{3/2}(z, T)}.\label{un}
\end{equation}
The pressure:
\begin{equation}
P=g\frac{(2\pi mT)^{3/2}T}{h^3}h_{5/2}(z, T) =\rho T \frac{h_{5/2}(z, T)}{h_{3/2}(z, T)}.\label{p}
\end{equation}
Knowing the energy per particle we can define the specific heat:
\begin{eqnarray}
\frac{C_V}{N}&=&  \frac{\partial U}{\partial T}\Big|_V\nonumber\\
&=&\frac{15}{4}\frac{h_{5/2}(z, T)}{h_{3/2}(z, T)}-\frac{9}{4}\frac{
C_{5/2}(z, \bar\varepsilon_H/T)+h_{3/2}(z, T)}{C_{3/2}(z, \bar\varepsilon_H/T)+h_{1/2}(z, T)}  \nonumber\\
&&+\frac{3}{2}\frac{\bar\varepsilon_H}{T} \frac{C_{5/2}(z, \bar\varepsilon_H/T)h_{1/2}(z, T)}{[C_{3/2}(z, \bar\varepsilon_H/T)+h_{1/2}(z, T)]h_{3/2}(z, T)}\nonumber\\
&&-\frac{3}{2}\frac{\bar\varepsilon_H}{T} \frac{C_{3/2}(z, \bar\varepsilon_H/T)}{C_{3/2}(z, \bar\varepsilon_H/T)+h_{1/2}(z, T)},\label{cv}
\end{eqnarray}
where 
\begin{equation}
C_n(z, x) = \frac{1}{\Gamma(n)}\frac{x^{n-1}}{z^{-1}e^{x}-1}.
\end{equation}
And incompressibility K
\begin{eqnarray}
K &=& 9\frac{\partial P}{\partial \rho}\nonumber\\
&=& 9 T\Big\{ \frac{C_{5/2}(z, \bar\varepsilon_H/T)+h_{3/2}(z, T)}{C_{3/2}(z, \bar\varepsilon_H/T)+h_{1/2}(z, T)} \nonumber\\
&&-\frac{2}{3}\frac{\bar\varepsilon_H}{T}\frac{ C_{5/2}(z, \bar\varepsilon_H/T)h_{1/2}(z, T)}{[C_{3/2}(z, \bar\varepsilon_H/T)+h_{1/2}(z, T)]h_{3/2}(z, T)}\nonumber\\
&&+\frac{2}{3}\frac{\bar\varepsilon_H}{T}\frac{ C_{3/2}(z, \bar\varepsilon_H/T)}{C_{3/2}(z, \bar\varepsilon_H/T)+h_{1/2}(z, T)}\Big\}.\label{incomp}
\end{eqnarray}
The factor 9 in Eq. (\ref{incomp}) is formally introduced in nuclear physics but ignored in other fields, notice that usually the compressibility is also defined  as the inverse of Eq. (\ref{incomp}). 
At zero T these quantities are given by
\begin{equation}
P_{H0} = \frac{2}{5}(\frac{3}{4\pi})^{\frac{2}{3}} \frac{\hbar^2}{2m} \rho^\frac{5}{3}, \hspace{0.1cm} K_{H0}=6(\frac{3}{4\pi})^{\frac{2}{3}}\frac{\hbar^2}{2m} \rho^\frac{2}{3}, \hspace{0.1cm} C_{V0}=0. \label{pt0}
\end{equation}
Notice the similarity again with a Fermi gas.  $\bar\varepsilon_H$, $P_{H0}$ and $K_{H0}$ are the natural units for our systems, while the temperature can be expressed in terms of the critical $T_0$,  and we will discuss the results  divided by such quantities.

In Fig. \ref{Fig1}, we plot the energy per particle, pressure and specific heat as function of temperature, all quantities are in adimensional form. The symbols represent the numerical solution to the BE gas with the modification due to the Heisenberg principle as discussed above. Three different densities are considered differing  orders of magnitude from each other. All different densities collapse in one single curve when scaled quantities are used.  The dashed line gives the result of the ideal BE gas \cite{landau, huang, pathria}. As we can see the two approaches give very similar results and start to deviate for temperatures approaching zero. In particular the energy per particle and pressure reach the values obtained from the Heisenberg principle, Eqs. (\ref{unl}) and (\ref{pt0}). The specific heat is well reproduced by the ideal case apart near $T=T_0$, where the discontinuity in the original case becomes a maximum. In particular the modified case displays a critical temperature slightly different from Eq. (\ref{ct}). We can easily understand this difference.  In fact for zero chemical potential, where the maximum is obtained, we get:
\begin{eqnarray}
\rho&=& \frac{g2\pi(2m)^{3/2}}{h^3}\int_{\bar\varepsilon_H}^\infty d\varepsilon \frac{\varepsilon^{1/2}}{e^{\varepsilon/T}-1} \nonumber\\
&=&\frac{g2\pi(2m)^{3/2}}{h^3}\Big[\int_{0}^\infty d\varepsilon \frac{\varepsilon^{1/2}}{e^{\varepsilon/T}-1}-\int_0^{\bar\varepsilon_H} d\varepsilon \frac{\varepsilon^{1/2}}{e^{\varepsilon/T}-1}\Big]\nonumber\\
&=&\frac{g2\pi(2mT)^{3/2}}{h^3}\Big[\int_{0}^\infty dx\frac{x^{1/2}}{e^{x}-1}-\int_0^{\bar\varepsilon_H/T} dx \frac{x^{1/2}}{e^{x}-1} \Big]\nonumber\\
&=& (\frac{T}{T_0})^{3/2}\Big[\rho-\frac{g2\pi(2mT_0)^{3/2}}{h^3}\int_0^{\bar\varepsilon_H/T} dx \frac{x^{1/2}}{e^{x}-1}\Big].
\end{eqnarray}
which explains the little shift observed in the figure. The presence of the maximum near $T_0$ suggests that we can still define a condensate of Bosons not at zero energy but at the average Heisenberg energy. However, as we see from the figure there is no phase transition at $T_0$ since all quantities are continuous, thus we deal more with a cross over rather than a first order phase transition such as in the ideal BE gas \cite{landau, huang, pathria}. Recall that in a real physical system, such as ${}^4He$, the specific heat displays a divergence at the critical temperature, which implies a second order phase transition. Furthermore the critical temperature is slightly less than estimated in Eq. (\ref{ct}). This is due to interactions among bosons which must be taken into account in order to reproduce the correct physical values \cite{he41, he42, he43, he44}.

An important quantity that is unphysical in the ideal BE gas case is the incompressibility below the critical point. In fact, since the particles that condense at zero energy give no contribution neither to the energy per particle or to the pressure below $T_0$, it turns out that the pressure is density or volume independent:
\begin{equation}
P = 1.3415\frac{g(2\pi mT)^{3/2}T}{h^3}. \label{pi}
\end{equation}
Thus its derivative Eq. (\ref{incomp}), which gives the incompressibility, is zero, an unphysical result. Within our framework the particle below $T_0$ 'condense' at the Heisenberg energy and pressure, which now depend on density similar to a Fermi gas.  In particular we expect that at zero temperature the incompressibility is different from zero and it is given by the Heisenberg value, Eq. (\ref{pt0}). At finite densities we can calculate the incompressibility by fixing the temperature and solving the modified BE integrals numerically. First we calculate the pressure for fixed temperature as function of density; the derivative of the pressure respect to density gives the incompressibility. 
\begin{figure}
\centering
\includegraphics[width=1.0\columnwidth]{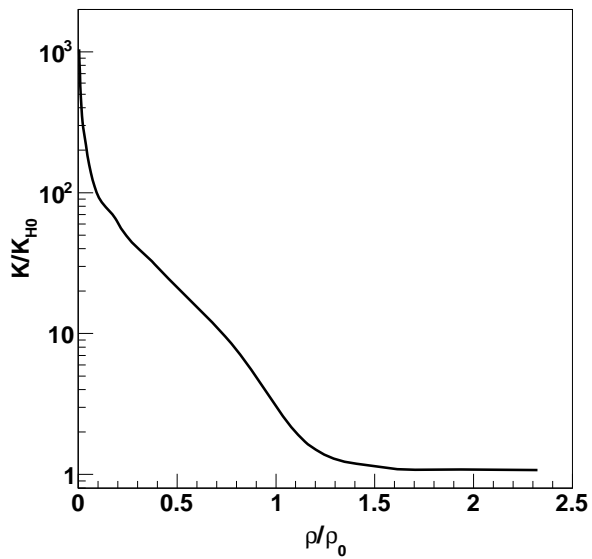}
\caption[]{Incompressibility as function of density for fixed $T$.} \label{Fig2}
\end{figure}
The incompressibility as function of reduced density is plotted in Fig. \ref{Fig2} for fixed $T$. Since the temperature is fixed, we expect that increasing the density until $T<T_0$ the incompressibility should reach the Heisenberg value given in Eq. (\ref{pt0}). Such a limit is indeed observed in the figure.  On the other hand, at very small densities, for fixed $T$, the pressure reaches the classical ideal gas value $P=\rho T$, see Eq. (\ref{p}). The incompressibility given by the derivative of the pressure respect to density, becomes a constant.  Since in Fig. \ref{Fig2} we divide the incompressibility by the Heisenberg incompressibility, in the limit $\rho \rightarrow 0$ we should get a divergence, exactly as shown.

In conclusion, in this brief report we have discussed a possible modification to the ideal BE gas to take into account the Heisenberg principle. We have shown that at zero temperature the Bosons get an average kinetic energy due to the indetermination principle. As a consequence the (in)compressibility is finite at all densities and temperatures in contrast to the ideal case. The specific heat has a maximum at a 'critical' temperature obtained at zero chemical potential. The position of the maximum is slightly shifted respect to the critical temperature of the ideal case. The jump in the specific heat is not observed, thus instead of a second order phase transition, the Heisenberg modification results in a cross-over to the condensate of particles which, we stress, has a finite kinetic energy even at zero temperature. Of course, interactions play an important role to reproduce the observations, nevertheless our modifications improve the quality of a non interacting Bose gas and take into account important physical properties such as the Heisenberg principle and finite (in)compressibility.  In the limit of zero densities the two approaches are identical as should be. The chemical potential can get positive values up to a maximum value given by the average Heisenberg energy, ideally reached in the limit of zero temperatures. Such a property is in contrast with the ideal case which dictates the chemical potential to be negative or zero at most.

\end{document}